\documentclass[aps,prl,twocolumn,groupedaddress]{revtex4}
\usepackage{graphicx}

\begin{document}


\title{ Determination of a Wave Function Functional }


\author{Xiao-Yin Pan}
\author{ Viraht Sahni}
\author{Lou Massa}

\affiliation{The Graduate School of the City University of New
York, New York, New York 10016. }




\begin{abstract}
 We propose expanding the space of
variations in traditional variational calculations for the energy
by considering the wave function $\psi$ to be a functional of a
set of functions $\chi: \psi = \psi[\chi]$, rather than a
function.  A constrained search in a subspace over all functions
$\chi$ such that the functional $\psi[\chi]$ satisfies a sum rule
or leads to a physical observable is then performed. An upper
bound to the energy is subsequently obtained by variational
minimization.  The \emph{rigorous} construction of such a
constrained-search--variational wave function functional is
demonstrated.

\end{abstract}


\maketitle

One of the mostly extensively employed and accurate
 approximation methods in
quantum mechanics is the variational principle for the energy.
Consider a quantum mechanical system with Hamiltonian operator
${\hat H}$. The ground state eigenenergies $E$ and eigenfunctions
$\Psi$ for this system satisfy the Schr{\"o}dinger equation $
{\hat H} \Psi=E\Psi$. Next define the functional $I[\psi] = \int
\psi^{*} {\hat H} \psi d\tau /\int \psi^{*} \psi d\tau  $.
Searching over all bounded, quadratically integrable functions
$\psi$, one obtains $I[\Psi] = E$, with $\Psi$ being the solution
of the Schr{\"o}dinger equation associated with energy $E$. Since
$\delta I[\Psi] = 0$, the functional $I[\psi]$ is stationary for
$\psi=\Psi$. In practice, an approximate function $\psi$ of a
particular analytical form is chosen to depend upon a number of
variable parameters $c_{i} ( i = 1,...,p)$. A least upper bound to
the ground state energy $E_{0}$ is obtained by solving the $p$
equations $\partial{I[\psi]} /\partial{c_{i}} = 0$,  and employing
the derived set of values of the parameters $c_{i}$ to calculate
$I[\psi]$ . In application of the variational principle, however,
the space of variations is limited by the choice of form of the
function chosen for the approximate wave function.  For example,
if Gaussian or Slater-type orbitals are employed,  then the
variational space is limited to such  functions. In this paper we
propose the idea of overcoming this limitation by expanding the
space over which the variations are performed. This then allows
for a greater flexibility for the structure of the approximate
wave function. We demonstrate the idea of expansion of
the variational space by example. \\

We  expand the space of variations by considering the approximate
wave function to be a functional of the set of functions $\chi$:
$\psi=\psi[\chi]$, rather than a function. The space of variations
is expanded because the functional $\psi[\chi]$ can be adjusted
through the function $\chi$ to reproduce any well behaved
function. However, this space of variations is  still too large
for practical purposes, and so we consider a \emph{subset} of this
space. In addition to the function $\psi$ being of a particular
analytical form and dependent on the variational parameters
$c_{i}$ , the functions $\chi$ are chosen such that the functional
$\psi[\chi]$ satisfies a constraint. Examples of such constraints
on the wave function functional $\psi[\chi]$ are  those of
normalization or the satisfaction of the Fermi-Coulomb hole charge
sum rule, or the requirement that it lead to observables such as
the electron density, the diamagnetic susceptibility, nuclear
magnetic constant or any other physical property of interest.  A
constrained-search over all functions $\chi$ such that
$\psi[\chi]$ satisfies a particular condition is then performed.
With the functional $ \psi[\chi]$ thus determined, the functional
$I[\psi[\chi]]$ is then minimized with respect to the parameters
$c_{i}$ . In this manner both a particular system property of
interest as well as the energy are obtained accurately, the latter
being a consequence of the variational principle.  We refer to
this way of determining an approximate wave function as the
\emph{constrained-search--variational} method.\\

As an example of the method we consider its application to the
ground state of the Helium atom. In atomic units $e=\hbar=m=1$,
 the non-relativistic Hamiltonian is
\begin{equation}
\hat{H}=(-1/2)\nabla_{1}^{2}-(1/2)\nabla_{2}^{2}-Z/r_{1}
  -Z/r_{2}+1/r_{12},
\end{equation}
where ${\bf r}_{1}$, ${\bf r}_{2}$ are the coordinates of the two
electrons, $r_{12}$ is the distance between them, $Z=2$ is the
atomic number. In terms of the Hylleraas coordinates\cite{1}
$s=r_{1}+r_{2}, \; t=r_{1}-r_{2}, \;  \; u=r_{12}$,  we choose the
ground state wave function functional to be of the general form
\begin{equation}
\psi[\chi]=\Phi(\alpha,s) [1-f(s,t,u)],
\end{equation}
with $\Phi(\alpha,s)=(\alpha^{3}/\pi)exp(-\alpha s)$,
$f(s,t,u)=e^{-q u}(1+qu)[1-\chi(q,s,t,u)(1+u/2)]$, where $\alpha$
and $q$ are variational parameters. \emph{Any} two-electron wave
function may be expressed in this most general
form. \\

In our example, we consider $\chi$ to be a function only of the
variable $s$: $\psi=\psi[\chi(q,s)]$. Then,  this wave function
satisfies the electron-electron cusp condition\cite{2}, which in
integral form is $\Psi({\bf r}_{1}, {\bf r}_{2},...,{\bf
r}_{N})|_{{\bf r}_{1}\rightarrow {\bf r}_{2}}=\Psi({\bf r}_{2},
{\bf r}_{2},...,{\bf r}_{N}) (1+\frac{r_{12}}{2})+{\bf
r}_{12}\cdot {\bf C}({\bf r}_{2}, {\bf r}_{3},...,{\bf r}_{N})$,
where ${\bf r}_{12}={\bf r}_{1}-{\bf r}_{2}$ and ${\bf C}({\bf
r}_{2}, {\bf r}_{3},...,{\bf r}_{N})$ is an undetermined vector.
This is readily seen  by Taylor expanding $\psi[\chi(q,s)]$ about
$u=0$. In a similar manner, the wave function $\psi[\chi(q,s)]$
satisfies the  electron-nucleus cusp coalescence
condition\cite{2}, which is  $\Psi({\bf r}, {\bf r}_{2},...,{\bf
r}_{N})|_{{\bf r}\rightarrow 0 }=\Psi(0, {\bf r}_{2},...,{\bf
r}_{N}) (1-Z r_{12})+{\bf r}\cdot {\bf A}({\bf r}_{2}, {\bf
r}_{3},...,{\bf r}_{N})$ with ${\bf A}({\bf r}_{2}, {\bf
r}_{3},...,{\bf r}_{N})$ being an undetermined vector, for
$\alpha=2$. For arbitrary value of the variational parameter
$\alpha$, the electron-nucleus coalescence condition is not
satisfied. (The differential form of these cusp conditions
obtained from the integral versions is $d \Psi_{sp.av}/dr
|_{r=0}=\zeta \Psi|_{r=0}$,  where $\Psi_{sp.av}$ is the spherical
average of the wave function about the singularity, ${\bf r}={\bf
r}_{1}-{\bf r}_{2}$,
 ${\bf r}_{1}$ and  ${\bf r}_{2}$  the positions of the two particles, and with $\zeta=1/2$ for electron-electron
 coalescence and $\zeta=-Z$ for electron-nucleus coalescence.)\\

The constraint to be applied to the wave function functional
$\psi[\chi]$ is the normalization condition:
\begin{equation}
\int \psi^{*}[\chi] \psi[\chi] d\tau=1.
\end{equation}
A pictorial representation of the  space of variation in the above
example relative to the standard variational method is given in
Fig.1. \\

The next step is the constrained search over functions $\chi(q,s)$
for which the condition of normalization is satisfied.
Substitution of our $\psi[\chi]$ into Eq.(3) leads to
\begin{equation}
2 \pi^2 \int_{0}^{\infty}ds |\Phi(\alpha,s)|^{2}g(s)=0,
\end{equation}
where
\begin{equation}
g(s)=\int_{0}^{s}u  du  \int_{0}^{u}dt (s^{2}-t^{2}) [f^{2}(s,u)-2
f(s,u)].
\end{equation}
Next consider Eq.(4). If the parameter $\alpha$ were fixed at say
the value for which the energy is minimized by the Hydrogenic
pre-factor $\Phi(\alpha,s)$, i.e. $\alpha=27/16$, then there exist
\emph{many} functions $g(s)$ for which the condition of Eq.(4) is
satisfied. These solutions correspond to the subspace C as shown
in Fig.1.\\

 On the other hand, if the parameter $\alpha$ is
variable, then the only way in which the condition of Eq.(4) can
be satisfied is if
\begin{equation}
g(s)=0.
\end{equation}
\emph{The requirement of Eq.(6) is equivalent to the constrained
search over the entire subspace C.}
\begin{figure}
 \begin{center}
 \includegraphics[bb=0 0 454 466, angle=359.5, scale=0.3]{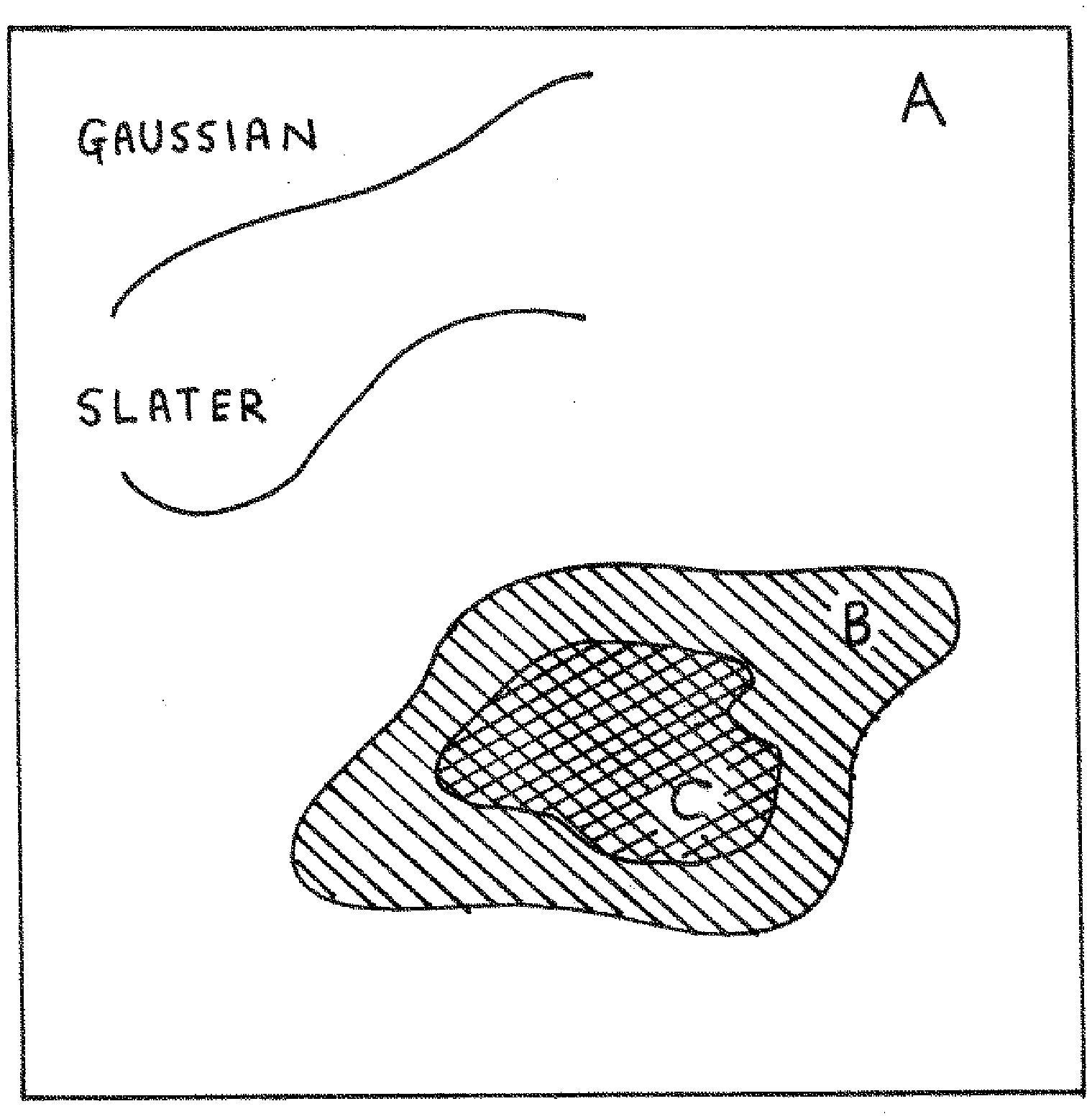}
 \caption{In the figure, the area of box A represents the space of
 all normalized square-integrable functions.  Functions
 such as Gaussian or Slater orbitals are represented by lines.
 The space A is also representative of the  functionals
 $\psi[\chi(s,t,u)]$. The subspace area B represents all
 normalized wave function functionals $\psi[\chi(s,t)]$, and its
 subspace C that of all normalized  functionals $\psi[\chi(s)]$
 .\label{}}
 \end{center}
 \end{figure}
Substitution of $f(s,u)$ into Eq.(6) in turn is equivalent to
 a quadratic equation for the function $\chi(q,s)$:
\begin{equation}
 a(q,s)\chi(q,s)^{2}+2 b(q,s) \chi(q,s)+c(q,s)=0,
\end{equation}
where
\begin{equation}
 a(q,s)=\int_{0}^{s} (s^{2}u^{2}-u^{4}/3)(1+u/2)^{2}(1+qu)^{2}e^{-2 qu}du,
\end{equation}
\begin{eqnarray}
 b(q,s)&=&-\int_{0}^{s}(s^{2}u^{2}-u^{4}/3)(1+u/2)(1+qu) \nonumber\\
     & &[e^{-2 qu}(1+qu)-e^{-qu}]du,
\end{eqnarray}
\begin{equation}
 c(q,s)=\int_{0}^{s} (s^{2}u^{2}-u^{4}/3)(1+qu)[e^{-2 qu}(1+qu)-2
 e^{-qu}]du.
\end{equation}
The integrals for the coefficients $a$, $b$, and $c$ are solvable
analytically. \emph{Solution of the quadratic equation is then
equivalent  to searching over the entire subspace C.}
\emph{Subspace C is comprised of only two points}. These two
points corresponding to the two solutions $\chi_{1}(q,s)$ and
$\chi_{2}(q,s)$ are such that the two wave functions
$\psi[\chi_{1}]$ and $\psi[\chi_{2}]$ are normalized. Rigorous
upper bounds to the ground state energy are then obtained by
variational minimization of the functional $I[\psi[\chi]]$ with
respect to the parameters $\alpha$ and $q$. The details of the
derivation and analytical expressions for the wave function
functionals are to be given elsewhere.
\\

The ground state energies obtained from the wave function
functionals $\psi[\chi_{1}]$ and $\psi[\chi_{2}]$ are given in
Table I together with those due to the energy-minimized Hydrogenic
pre-factor $\Phi(s)$ with $\alpha=27/16$, the Hartree-Fock(HF)
 value\cite{3}, and the `exact' value of Pekeris\cite{4}.
The functions $\chi_{1}(q,s)$ and $\chi_{2}(q,s)$ at the
respective energy minimum  are plotted in Fig.2. We also quote
energy minimized values determined from the wave function
functionals $\psi[\chi_{3}]$ and $\psi[\chi_{4}]$
 obtained by solution of the quadratic equation with $\alpha$
 fixed at $\alpha=27/16$, and $q$ treated as the only variational
 parameter. (The functions ($\chi_{3}$, $\chi_{4}$) differ
 minimally from those of ($\chi_{1}$, $\chi_{2}$).) The
 satisfaction of the virial theorem and the expectation value of the operator $W=r_{1}+{r}_{2}$ are also quoted.
   \\

We note the following  points of interest: (i) The functions
$\chi_{1}(q,s)$ , $\chi_{2}(q,s)$ are very different:
$\chi_{1}(q,s)$ is a positive monotonically decreasing function
whereas $\chi_{2}(q,s)$ is a negative monotonically increasing
function. They, however, are of about the same magnitude. Thus,
the  functionals $\psi[\chi_{1}]$ and $\psi[\chi_{2}]$ are
different from each other. Similarly, the functionals
$\psi[\chi_{3}]$ and $\psi[\chi_{4}]$ differ.(ii) In spite of the
wave function functionals $\psi[\chi_{1}(q,s)]$ and
$\psi[\chi_{2}(q,s)]$ being different, the corresponding energies
are essentially equivalent. (iii) Note that
$E[\chi_{2}]<E[\chi_{1}]<E[\chi_{4}]<E[\chi_{3}]$. (iv) The
improvement in the energies noted in (iii) is also reflected in
the corresponding satisfaction of the virial theorem.
  (v) As the constrained search is
over the normalized  functionals of subspace C,
 \emph{both} solutions of the quadratic
equation Eq.(7) lead to meaningful energies.\\

Our results  demonstrate the advantage of the concept of a wave
function functional. (The purpose of the paper is not to determine
the most accurate wave function for the He atom ground state.) The
energies obtained via the various wave function functionals are to
our knowledge the most accurate one- and two-parameter results in
the literature. A $1.93\%$ error of the Hydrogenic pre-factor is
reduced to errors of $0.45\%-0.43\%$ for the four wave function
functionals. The corresponding satisfaction of the virial theorem
is $1.61\%-0.08\%$. As a further point of comparison, we note that
our results are superior to those of HF, and the results of
$\psi[\chi_{1}]$, $\psi[\chi_{2}]$ and $\psi[\chi_{4}]$ are also
superior to a $3$-parameter wave function calculation(CK)
\cite{5}. ( This wave function is of the form of Eq.(2) with
$f(s,u)=e^{-qu}(1+qu)[1-\beta e^{-2\mu s}(1+u/2)]$ with
$\alpha=27/16$, $q=0.562326$, $\mu=0.099947$, $\beta=0.88066$).
The exact satisfaction of the virial condition by the pre-factor
is a consequence of scaling, and that of HF due to the
self-consistent nature of the solution. That of Pekeris is a
consequence of the accuracy of the wave function. The expectation
values of  $W=r_{1}+r_{2}$ are of course only expected to be as
accurate as that of wave function, and not correct to second order
as is the energy. The values obtained by the functionals
$\psi[\chi_{1}]$ and $\psi[\chi_{2}]$ are an improvement over
those of the pre-factor, reducing an error of $4.3\%$ to $2.9 \%$.
They are also significantly superior to those obtained by the
$3$-parameter wave function which is in error by $3.9 \%$. Similar
improvements are observed for the expectations of other single
particle operators. On the other hand, the high accuracy of the HF
value is because in this theory single-particle
 expectation values are correct to second order \cite{6,9}. Our results also
 demonstrate that by expanding the space of variations, more accurate results
 for both the energy as well as other properties can be obtained with fewer variational parameters.  \\

An improvement over the present results can be achieved as
follows. (i) Expand the space of variations by considering $\chi$
to be a function of the variables $s$ and $t$, or expand the space
still further by considering $\chi$ to be a function of the
variables $s, t$, and $u$. Such an expansion of the space could
require the solution of an integral equation for the determination
of $\chi$. (ii) Replace the Hydrogenic pre-factor $\Phi(\alpha,s)$
by the analytical HF wave function. (iii) Combine the expansion of
the variational space with the improvement of the
pre-factor. (iv) Employ a different analytical form for the wave function functional.\\

As noted above, it is also possible to search over all functions
$\chi(q,s)$ such that the functional $\psi[\chi]$ leads to a
physical property of interest. For example, let us consider the
expectation of  $W=r_{1}+r_{2}$: $<r_{1}+r_{2}>=<s>=s_{0}+\Delta s
$. Here $s_{0}$ is the expectation from the pre-factor $\Phi(s)$.
Assuming $\Delta s$ known from experiment or some accurate
calculation, and if a wave function functional $\psi[\chi]$ of the
form in the above calculation is employed, then two distinct
$\chi$'s such that $<\psi[\chi]|s|\psi[\chi]>=s_{0}+\Delta s$ can
be obtained by solution of the quadratic equation
$a(q,s)\chi(q,s)^{2}+2 b(q,s)\chi(q,s) +[c(q,s)-A]=0$, where the
constant $A=2 \Delta s/\alpha^{4}$, and where the coefficients
$a,b$ and $c$ are the same as in Eq.(8-10). With the functionals
$\psi[\chi_{1}]$ and $\psi[\chi_{2}]$ thus determined, the energy
could then be obtained by minimization of the functional
$I[\psi[\chi] ]$ with respect to the parameters $\alpha$ and $q$.
In this manner, the two wave function functionals would reproduce
both the size of the atom exactly \emph{and} the energy
accurately. \\

\begin{figure}
 \begin{center}
 \includegraphics[bb=0 0 512 722, angle=90, scale=0.3]{figure2.eps}
 \caption{The functions $\chi_{1}(q,s)$ and $\chi_{2}(q,s)$
 .\label{}}
 \end{center}
 \end{figure}

For completeness we note that the concept of constrained search in
the present work differs from that\cite{7} within density
functional theory (DFT). The  idea underlying DFT is based on the
first Hohenberg-Kohn theorem\cite{8} according to which the wave
function $\Psi$ is a functional of the \emph{ground} state density
$\rho({\bf r}): \Psi = \Psi[\rho]$.  Thus the energy is a unique
functional of $\rho({\bf r})$ : $E = E[\rho]$. The \emph{in
principle} constrained search  within DFT is as follows. One first
searches for the infimum of the $\langle{\hat H}\rangle$ over all
antisymmetric , normalized, $N$-particle functions $\Psi$ whose
density is $\rho({\bf r})$. One then searches over all such
$\rho({\bf r})$ to obtain the infimum of that expectation. The
consecutive infima can be shown to be a minimum and equal to the
ground state energy. In our work, the statement that the wave
function is  a functional of the functions $\chi$ is more general.
The functions $\chi$ are not restricted to just being the density
or functions of the density. And the constrained search over all
functions $\chi$ is such that the wave function leads to an
\emph{arbitrary} property of interest. The
energy is subsequently obtained by variational minimization.\\

In addition to its use in Schr{\"o}dinger theory, the wave
function functional $\psi[\chi]$ may also be employed within
Quantal density functional theory (Q-DFT)\cite{9}.  In Q-DFT,  as
in traditional Kohn-Sham (KS) DFT\cite{10}, the system of
electrons is transformed into one of \emph{noninteracting}
fermions such that the same ground state density, energy and
ionization potential are obtained.  However, in contrast to KS-DFT
which  is in terms of energy functionals and functional
derivatives, the Q-DFT framework involves fields whose quantal
sources are expectations of Hermitian operators taken with respect
to the wave function. Thus, an approximate wave function
functional of the form of Eq.(2) can be employed in this theory
with the Slater determinantal pre-factor  determined
\emph{self-consistently}. The wave function functional $\psi[\chi
= f(\rho)]$ could also be used. Within Q-DFT, the corresponding
energy
is a rigorous upper bound.\\

\begin{table}
\caption{\label{tab:table I} Rigorous upper bounds to the ground
state energy of the Helium atom in (a.u) for different wave
functions (WF). The satisfaction of the  virial theorem, and the
expectation values of $W=r_{1}+r_{2}$ are also quoted.}
\renewcommand{\arraystretch}{0.1}
\begin{tabular}{|c |c |c |c|c|}
\hline \hline WF  & Parameters &  Energy & $-V/T$ & $< W
>$  \\ \hline
$\Phi$ & $\alpha=1.6875$ & $-2.84766$& $2.0000$ & $1.7778$ \\
\hline \hline

 $\psi[\chi_{3}]$& $\alpha=1.6875$ $q=0.581$ & $-2.89004$&1.9678 &$1.7778$ \\ \hline
 $\psi[\chi_{4}]$& $\alpha=1.6875$ $q=0.180$ & $-2.89061$&1.9707 &$1.7778$\\ \hline
$\psi[\chi_{1}]$& $\alpha=1.6614$ $q=0.5333$ & $-2.89072$&1.9973 &
$1.8057$
\\\hline $\psi[\chi_{2}]$& $\alpha=1.6629$ $q=0.17049$ &
$-2.89122$ &1.9984 & $1.8041$\\\hline \hline
 $HF$ & See Ref. 3  & $-2.86168$& $2.0000$ & $1.8545$\\ \hline
$CK$ &See Ref. 5 & $-2.89007$& 1.9890 & $1.7848$\\
\hline
$Pekeris$ &See Ref. 4 & $-2.90372$& $2.0000$ & $1.8589$\\
\hline \hline
\end{tabular}
\end{table}

We also mention the work of Colle and Salvetti(CS) \cite{11} who
suggested a wave function functional of the density. The CS wave
function is similar to Eq.(2) except that the pre-factor is the HF
wave function and the correlation term is $f({\bf r}_{1},{\bf
r}_{2})=e^{-\beta^{2} r^{2}}[1-\chi({\bf R})(1+r/2)]$, with ${\bf
r}={\bf r}_{1}-{\bf r}_{2}$, ${\bf R}={\bf r}_{1}+{\bf r}_{2}$,
$\beta=q [\rho^{HF}({\bf R})]^{1/3}$. CS further assumed that the
corresponding Dirac density matrix was that due to the pre-factor.
The function $\chi({\bf R})$ was to be determined by requiring
that the correction to the HF Dirac density matrix due to the
correlation factor $f({\bf r}_{1},{\bf r}_{2})$ vanishes. Since
the HF Dirac density matrix cannot be the same as the exact
interacting system density matrix, this is not an exact
constraint. Moreover, they did not satisfy this condition, and
instead approximated the function $\chi({\bf R})$ by $\chi({\bf
R})=\sqrt{\pi} \beta/(1+\sqrt{\pi} \beta$). A consequence of this
was that the resulting wave function was not
normalized\cite{12,5,13}.  There is also no discussion in this
work of the general concept of wave
function functionals or the idea of constrained search to obtain them.\\

In conclusion, we have proposed the idea of expanding the space of
variations beyond that of standard variational calculations by
considering the wave function to be a functional  instead of a
function, a functional of the functions $\chi$.  A
\emph{constrained search} is performed over the functions $\chi$
such that the wave function satisfies a constraint or leads to a
physical observable. A rigorous upper bound to the energy is then
obtained by variational minimization with respect to any
parameters in the wave function functional.  The construction of
such a constrained-search--variational wave function functional
for the ground state of the Helium atom where the search is over
the \emph{entire} requisite  subspace is explicitly demonstrated.  \\

\begin{acknowledgments}
This work was supported in part by the Research Foundation of
 CUNY. L. M. was supported in part by NSF through CREST, and by
 a ``Research Centers in Minority Institutions'' award, RR-03037,
 from the National Center for Research Resources, National
 Institutes of Health.
\end{acknowledgments}

\end{document}